# Analyzing Turkish F and Turkish E keyboard layouts using learning curves

*Mevlut Serkan Tok[*,a], Osman Tufan Tekin[a], Kemal Bicakci[b]*

[a]*Computer Engineering Department, Graduate School of Engineering and Science, TOBB University of Economics and Technology, Ankara, Turkey*
[b]*Computer Engineering Department, TOBB University of Economics and Technology, Ankara, Turkey*

A B S T R A C T

The F-layout was introduced in 1955 and eventually enforced as a national standard as a replacement to the popular QWERTY keyboard layout in Turkey. In a more recent work, another alternative (E-layout) was developed for Turkish language and argued to be faster and more comfortable than the F-layout. However, there has not been any empirical evidence favouring any of these layouts so far. To fill this research gap in the literature, we have employed a hybrid model and conducted both between-subjects and within-subjects user experiments with twelve freshmen majoring in computer engineering. The experimental results show that there is no significant difference between learning percentage of these two layouts but the completion time of typing a trial passage with the F-layout is significantly lower than the E-layout. The F-layout has also a significantly lower physical demand score, as revealed by the subjective assessments of participants. Based on our user survey data, we also discuss some possible reasons of F-keyboard's limited prevalence among Turkish users.

## 1. Introduction

QWERTY keyboard layout has de facto become the world's standard keyboard layout. Over the years, many efforts of designing an optimized, more efficient layout have been pursued by researchers. One of the most distinguished researchers in this area is August Dvorak. His famous simplified Dvorak layout was introduced in 1936 [1], but it has failed to replace the QWERTY despite its claimed advantages including reducing errors and requiring less finger motions [2].

After the Turkish alphabet reform in 1928, studies for designing a new keyboard were also launched in Turkey. With the distinctive contribution of Ihsan Sitki Yener, the Turkish F-layout design was introduced and proclaimed to be the standard Turkish keyboard layout in 1955 [3]. By contrast with arguments emphasizing re-arrangement of letters on keyboards brings negligible improvements [4], F keyboard got 17 world records in stenography competitions [5]. However, despite all the efforts to make the F-layout mandatory in the public sector [6], a QWERTY-layout keyboard, modified to include the Turkish-specific characters, is currently the most dominant keyboard in Turkey [7].

In 2014, another layout, which is totally different from the F-layout, but also optimized specifically for Turkish language, was designed [8]. Although the top left letter row starts with Q, not E, it is called the E-layout in order to avoid confusion with the QWERTY layout. The inventors of E-layout discussed many methods and technologies to use in developing their patented keyboard layout. For instance, they used motion capture gloves to measure tendon movements. They also argued that the E-layout is superior to the F-layout with respect to typing speed, ergonomics and user comfort level [9].

The analysis and comparison of these two alternatives remain an open problem for the academic literature. Up to our best knowledge, our work is the first study which aims at comparing the Turkish F-layout and E-layout through a formal user-based testing. In more general terms, our work contributes to the knowledge on alphabetical keyboard layouts with the findings of a well-documented and repeatable human experiment. The survey we conducted before and after the experiment also reveals interesting findings regarding the reasons why the F-keyboard has limited prevalence among Turkish users.

The rest of the paper is organized as follows: literature review is given in Section 2, the methodology is explained in Section 3, the results are presented in Section 4, a discussion is presented in Section 5, and our conclusions are presented in Section 6.

## 2. Background

*2.1. Learning Curve*

The learning curve was first used for describing the effects of learning in the aircraft industry and several standard equations were presented to calculate the learning percentage [10]. When an individual or an organization engages in doing something repeatedly, the time or cost required to do it decreases and eventually reaches a stable level as the number of repetitions increases because the individual's or organization's proficiency and experience continuously improve in the meantime [11].

$$N = \log(Y_x/K)/\log(X) \quad (1)$$

$$\text{Learning \%} = 100 * 2^N \quad (2)$$

Where $Y_x$ is the production time of the Xth unit in the sequence, K is the time required for producing the first unit, X is the number of production units, N is the slope of the line describing the change in completion time (y) as a function of the repetition number (x) in log-log space. Using (1) and (2), we calculated learning percentage of keyboard layouts in our study.

*2.2. Visual Analog Scale (VAS)*

Visual analogue scale is a measurement scale that is used to measure a characteristic or attitude that is believed to range across a continuum of values and cannot be easily measured [12]. It has been used in medical research as a pain rating scale for many years [13-15]. Visual analogue scales are also used to measure the level of mental and physical demand of workload on subjects [16].

*2.3. Using the learning percentage and visual analog scale to analyze keyboards*

A study conducted in 2009 shows that productivity decrements can be quickly regained for the QWERTY based split fixed-angle and contour split keyboard but will take considerably longer for Dvorak and chord keyboards. The split fixed-angle keyboard involved physical learning, whereas the others involved some combination of physical and cognitive learning. Multiple trials were performed while testing the keyboards in two different protocols to determine the learning percentages. Subjective assessment data was collected by using visual analogue scales to rate physical demand, cognitive demand and perceptual demand [17]. The experimental methodology of the study has been tailored, modified, and implemented in our research.

## 3. Methods

*3.1. Participants*

Twelve participants were recruited from the Department of Computer Engineering, TOBB University of Economics and Technology, all of whom were freshman students. The participants ranged in age from 19 to 21, with an average age of 19.6. All participants were right handed with 20/20 or corrected to 20/20 vision. A pre-experiment survey was implemented on participants to gather demographic data and to find out if they had current or chronic back, shoulder, neck, wrist or finger pain or if they typed with a TR-F and TR-E keyboard layout before. Potential participants were excluded from the study if they had any of these health problems or previous typing experience on alternative keyboard layouts. The participants were required to be able to type at least 25 words per minute (wpm) on the TR-Q keyboard [18]. The average typing speed of the participants on the TR-Q keyboard was 49.5 wpm (standard deviation 12.46; range 34 to 71) and all used a TR-Q keyboard regularly.

Two groups, protocol-1 and protocol-2, were established based on the average typing speed of whole participants. Detailed demographic data belong to participants is given in Section 4.5.

In terms of improving experiment and survey methodologies, four additional graduate students were employed to participate pilot study.

*3.2. Equipment*

The participants were asked to key on three distinctive keyboard layouts: TR-Q keyboard (Everest/KB-250U) shown in Fig. 1., TR-F keyboard (Everest/KB-250F) shown in Fig. 2., and TR-E keyboard (Everest KB-250U) shown in Fig. 3.

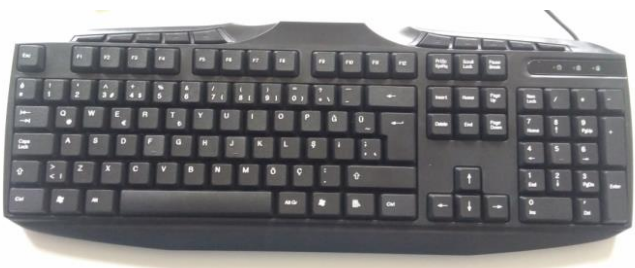

**Fig. 1.** TR-Q keyboard used during research.

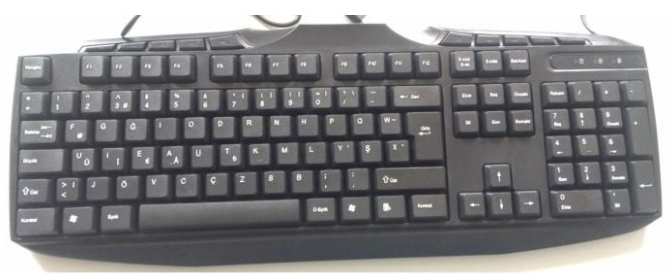

**Fig. 2.** TR-F keyboard used during research.

The typing trials were performed on a 15.6-inch laptop with Windows 10 installed (1366x768 resolution) and freeware typing program Stamina 2.0. Since Stamina 2.0 does not support Turkish characters and keyboard layouts, the file Layouts.ini was modified. A 2-min educational video was prepared to introduce the software to the participants in a standardized way. In Stamina 2.0, text moves across the screen line by line as shown in Fig. 4. No forward move is permitted when a wrong key is pressed. As Stamina 2.0 compels to type passage correctly, it hinders the typist from making a speed-accuracy trade-off by holding accuracy fixed at perfection while completion time diversified.

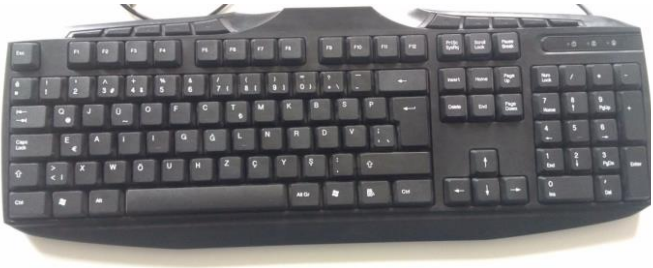

**Fig. 3.** TR-E keyboard used during research.

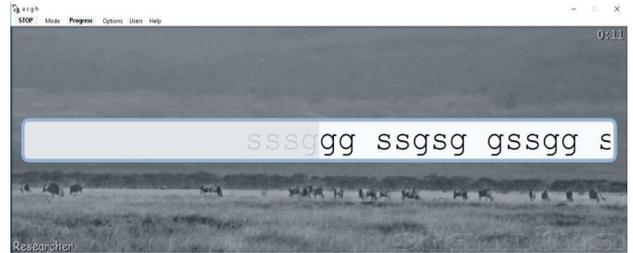

**Fig. 4.** Stamina 2.0 freeware typing program screenshot.

*3.3. Experimental Design*

In this study, a repeated-measures experimental design with one independent variable (keyboard) with two levels (TR-F and TR-E keyboards) was employed. The dependent variables in this study were learning percentage and subjective assessment of cognitive demand, physical demand and perceptual demand.

The trial passage was prepared in Turkish, contained 225 characters including spaces, contained all letters of Turkish alphabet in concordance with Turkish language common frequency-of-use data [19], two commas, three periods, and no numbers. After each trial, the experimenter recorded completion time in seconds, error percentage, and typing speed (character per minute).

To collect subjective assessment data, participants asked to rate each alternative keyboard in comparison to the TR-Q keyboard in terms of three workload demands, including physical demand (finger coordination, awkward postures, muscle force or tension, awkward reaches with fingers), cognitive demand (remembering, thinking, deciding, or planning) and perceptual demand (looking, searching, detecting, or recognizing) in the subjective assessment test [17].

The survey was based on the form of the NASA Task Load Index questionnaire [20]. Since the SI system is used in Turkey, the metric units were used in the experiment, rather than inches. Each VAS used in the demand rating survey was 10 cm in length with low and high anchors as well as mid-line drawn at 5 cm, identified on the VAS as the TR-Q keyboard condition. Defining the scale mid-line in terms of the TR-Q allowed for the other keyboard layout designs to be rated above or below it in terms of workload for all three demand types. Any keyboard scoring above 5 cm on the VAS was rated higher in demand than the TR-Q keyboard, and any keyboard scoring lower than 5 cm was rated lower in demand than the TR-Q keyboard.

*3.4. Pilot Study*

Four graduate students participated in the pilot study. Two of them took part in Protocol-1 and the others took part in Protocol-2. While experimenter-1 conducted a protocol with each participant, experimenter-2 observed the participant and took notes regarding surveys, trials, participant-keyboard and participant-computer interaction. After each trial, experimenter-1 recorded the completion time in seconds, error percentage, and typing speed (character per minute), but this data was not taken in to account. An interview was made after trials with each participant in the light of notes taken during their observation, and the experimental design was modified based on the findings of the pilot study.

*3.5. Protocol-1*

Eight participants participated in this protocol to determine their learning percentages for the alternative keyboards. Before the typing trials, each participant was given 1 minute to review the three-sentence trial text. After reviewing the passage, the participant was asked to type the passage 10 times on the TR-Q keyboard and to type as quickly as possible to record a baseline QWERTY typing speed. 15-second rest was given to each participant between successive typing trials.

After finishing 10 trials on the TR-Q keyboard, the participant was given a cheat sheet showing location of letters on the TR-F keyboard and TR-E keyboard. Then, the participant typed the same passage five times on the TR-E keyboard and five times on the TR-F keyboard. The order of keyboard presentation was randomized, and all keyboards were equally presented. The same passage was used in each trial. After each of five trials, the participant took a 15-sec break. With the help of this experimental design, a learn-by-doing process was followed.

After a set of five trials with each alternative keyboard, subjective assessment test was taken by the participant. Then a 3-minute break was given.

*3.6. Protocol-2*

Four participants were asked to participate in experimental protocol-2 in order to validate the use of the five-trial protocol for establishing estimates of the learning percentages for the keyboards. In this protocol, each participant performed 10 trials on the TR-Q keyboard. Then, the participants were divided into two groups based on their typing speed on the TR-Q keyboard (to establish two similar groups for the trials on the TR-F and TR-E keyboards): two participants performed 10 trials on the TR-F and the others performed 10 trials on the TR-E. The resting times between trials were arranged on the same terms as protocol-1. No subjective assessment test was implemented in protocol-2.

*3.7. User Surveys*

Two surveys were prepared: a pre-experiment survey and a post-experiment survey. All twelve participants attended these surveys. The pre-experiment survey had three goals: to collect data on age and gender, to clarify whether the participants had any medical problems or not, and to determine participants' habits regarding computer use and their prejudices about alternative keyboard layouts. Fifteen questions were asked. Eleven of them were statements and a Likert scale was used for obtaining answers while four questions were varied. Some questions were paraphrased into negative statements and asked again to provide justification. Invalidated results were not taken into consideration. The pre-experiment survey was implemented before the protocols.

The post-experiment survey had only one goal. Supposing that after multiple trials, the participants became familiar with alternative keyboard layouts, the post-experiment survey was conducted to figure out if the participants' prejudices about alternative keyboards had changed. Twelve statements were presented to the participants and respective answers obtained via a Likert scale. Some questions were paraphrased into negative statements and asked again to provide justification. Invalidated results were not taken into consideration. The post-experiment survey was implemented after the protocols.

In neither of the surveys, a comparison of the TR-F and TR-E keyboards was done.

*3.8. Data Processing and Statistical Analysis*

To calculate the learning percentage, equations (1) and (2) were used for the data obtained from protocol-1. The completion time for trial-1 and completion time for trial-5 were used to calculate the learning percentage. The data obtained from protocol-2 was processed on the same terms. In that case, the completion time for trial-1 and completion time for trial-10 were used.

Subjective ratings of participants regarding alternative keyboard demands were measured from the mid-lines of the VASes. A score ranging from -40 to 40 was given for each demand assessment (each point representing 1.25 mm from a scale mid-line). Survey scores were then normalized to reduce inter-participant variability [16]. For each demand category, paired t tests were performed to determine significant differences. A p value less than 0.05 was the standard for significance.

Since only two keyboards were tested in our study, paired t-test was employed to evaluate the effect of keyboard on learning in protocol-1. A p value less than 0.05 was the standard for significance. 8 participants conducted protocol-1 (5 trials), so the Shapiro-Wilk test of normality was used on the data obtained from the 5 trial tests to make sure the data was in the form of normal distribution.

An independent t-test was performed to evaluate the differences between the learning percentages after protocol-1 (5 trials) and the learning percentages after protocol-2 (10 trials). A p value less than 0.05 was the standard for significance.

As one independent variable was tested (keyboard) with two levels (the TR-F and TR-E keyboards), a paired t-test was performed to evaluate the differences between the trial completion times of alternative keyboards in protocol-1 (5 trials). A p value less than 0.05 was the standard for significance.

To determine whether the typing ability on the QWERTY keyboard affects the typing performance on alternative keyboards, the timings of the three fastest QWERTY typists, among those who did protocol 1, on alternative keyboards were compared to the timings of the three slowest QWERTY typists on alternative keyboards through an independent t-test. A p value less than 0.05 was the standard for significance.

A Pearson's bivariate correlation analysis was performed to evaluate the correlation between subjective levels of cognitive, physical and perceptual demand and learning percentages. A paired t-test was performed to evaluate the differences between the keyboards in terms of subjective levels.

All the statistical analyses were performed with IBM SPSS Statistics V.22.

## 4. Results

*4.1. Learning Percentages*

The learning percentage data sets for the TR-E (avg. 86.94%) and TR-F (avg. 88.33%) keyboards in protocol-1 (5 trials) were compared by means of the paired t-test: p=0.259 (p>0.05) and no significant difference was observed.

The learning percentages of protocol-1 (5 trials) and protocol-2 (10 trials) were compared with the independent t-test for each keyboard, as shown in Table 1., and no significant difference was observed.

**Table 1. Comparison of the learning percentages obtained from protocol-1 and protocol-2**

| Keyboard Type | Learning % | | | | |
| --- | --- | --- | --- | --- | --- |
| | Protocol-1 | | Protocol-2 | | |
| | M | SD | M | SD | P Value |
| TR-F | 88.33 | 4.03 | 85.65 | 3.72 | 0.47 |
| *TR-E* | 86.94 | 1.92 | 86.09 | 0.25 | 0.48 |

*4.2. Completion Time*

The completion time data sets for the TR-E and TR-F keyboards in protocol-1 were compared by means of the paired t test: p=0.001 (p<0.05), indicating that there is a significant difference between the TR-F and TR-E keyboards in respect of the completion times. The average timing results for the keyboards during protocol-1 are shown in Fig. 5. The TR-Q timings are only included to give an opinion regarding baselines.

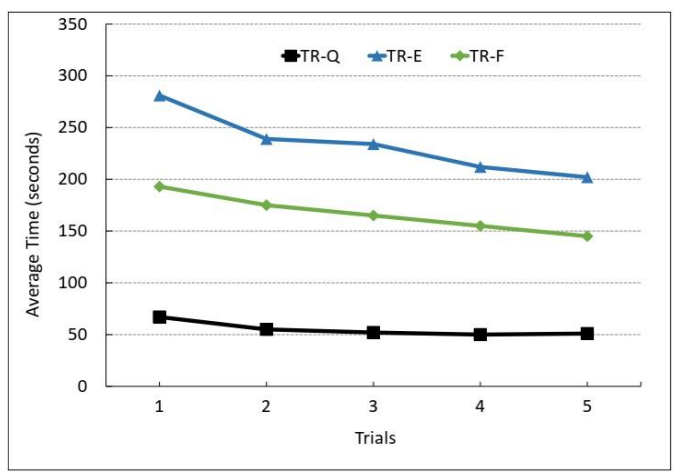

**Fig. 5.** Completion time of trials in protocol-1.

The normalized completion time of trials for each keyboard is shown in Fig. 6. in order to indicate how the time to complete is reduced with repetition. It can be seen the time reduction for the TR-E keyboard is more than that for the TR-F keyboard, which is consistent with the lower learning percentage associated with the TR-E keyboard.

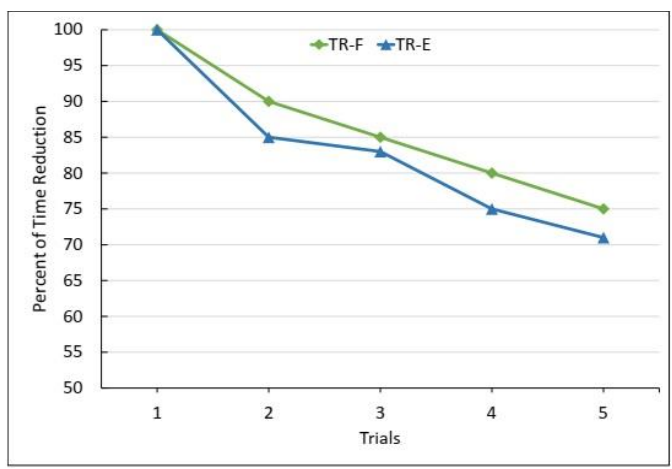

**Fig. 6.** Normalized completion time of trials in protocol-1.

To compare the participants' performance on alternative keyboards based on their typing ability on the TR-Q keyboard, an independent t-test (p<0.05) was conducted, which yielded (p= 0.053>0.05) to the conclusion that there is no significant difference between the completion times of the three fastest TR-Q typists (avg. 57 wpm) and three slowest TR-Q typists (avg. 35 wpm) achieved in protocol-1.

*4.3. Subjective Assessment of Demands*

As shown in Table 2. negative correlations were found for each aspect examined during the subjective assessment of demands, but only perceptual demand had a significant and strong correlation related to the learning percentage.

**Table 2. Correlations between learning percentage and demand type**

| Demand Type | Correlation Coefficient | p Value |
|---|---|---|
| Cognitive Demand | -0.35 | 0.182 |
| Physical Demand | -0.16 | 0.554 |
| Perceptual Demand | -0.53 | 0.034 |

In all three demand categories, as shown in Fig. 7., TR-E has the higher demand rates, but only physical demand rate has significant difference statistically as shown in Table 3.

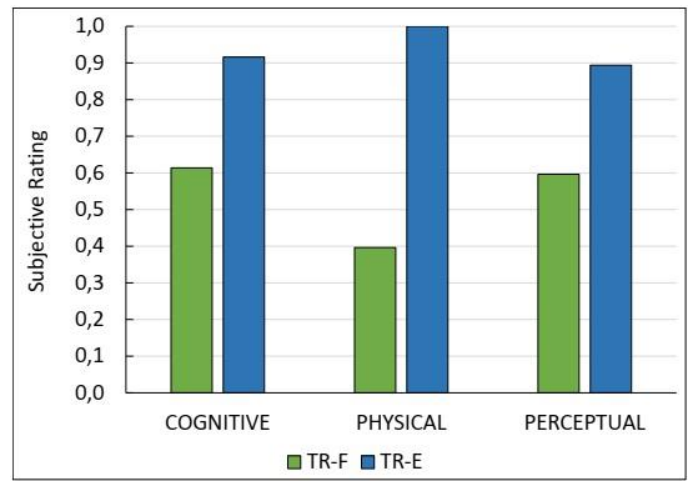

**Fig. 7.** Subjective assessment of demands as a function of the keyboard type.

**Table 3. Paired t-test results in terms of keyboards' demand rates**

| Demand Type | p Value |
|---|---|
| Cognitive Demand | 0.082 |
| Physical Demand | 0.018 |
| Perceptual Demand | 0.260 |

*4.4. Pre-Experiment Survey*

It was reported that most of participants (66.7%) used 6 to 8 fingers while typing on a TR-Q keyboard. Half of the participants stated that they typed both in Turkish and in English equally. Most of participants (83.3%) spent four hours or more on computer every day. For most of the participants, their main activities on computer was playing computer games as shown in Table 4.

**Table 4. Pre-experiment survey results (demography and computer usage habits)**

| Demographic and Computer Usage Habits | | N | % |
|---|---|---|---|
| Age | 18-19 | 6 | 50 |
| | 20-21 | 6 | 50 |
| Sex | Male | 5 | 42.7 |
| | Female | 7 | 58.3 |
| Typing with finger | 2-4 | 1 | 8.3 |
| | 6-8 | 8 | 66.7 |
| | 10 | 3 | 25 |
| Typing language | mainly Turkish | 5 | 42.7 |
| | mainly English | 1 | 8.3 |
| | TR-ENG equally | 6 | 50 |
| Hours spent on PC | 2-4 | 2 | 16.7 |
| | 4-6 | 7 | 58.3 |
| | >6 | 3 | 25 |
| Main activities on PC (Multiple choices allowed) | Homework/Research | 5 | 42.7 |
| | Gaming | 8 | 66.7 |
| | Stream (movies etc.) | 4 | 33.3 |
| | Social Media | 2 | 16.7 |

According to the pre-experiment survey results, all participants were informed about at least one alternative keyboard layout. Only half of them thought that typing fast on alternative keyboard layouts would contribute to their career and none of them wanted to own a laptop with an alternative keyboard layout. Only 16% of participants agreed that it was necessary to type on alternative layout keyboards and 41% of participants agreed that they did not have any prejudices regarding alternative keyboard layouts as shown in Table 5.

**Table 5. Pre-experiment survey results (opinions regarding alternative keyboard layouts)**

| Statement | Percentage of Agreement (Agree/Strongly Agree) % |
|---|---|
| I want to have a laptop with an alternative keyboard layout. | 0 |
| I think it is necessary to type on alternative keyboard layouts. | 16 |
| I have no prejudices regarding alternative keyboard layouts. | 41 |
| I believe typing fast on alternative keyboard layouts will contribute to my professional career. | 50 |
| I heard about alternative keyboard layouts. | 100 |

*4.5. Post-Experiment Survey*

According to the post-experiment survey results, after the experiment there is a slight increase in the percentage of participants who want to have a laptop with an alternative keyboard layout (8%). 41% of participants agree that it is necessary to type on alternative keyboard layouts and 66% of them agree that they believe typing fast on alternative layout keyboards would contribute to their professional career. 66% of participants agree that they want to type with ten fingers on an alternative keyboard layout. 92% of participants agree that it is possible to type fast on alternative keyboard layouts and 66% of participants agree that they do not have any prejudice regarding alternative keyboard layouts as shown in Table 6. The statements used in both the pre-experiment and post-experiment surveys were shown in the same order on Table 5. and Table 6. to emphasize how the participants' attitude has changed.

Table 6. Post-experiment survey results

| Statement | Percentage of Agreement (Agree/Strongly Agree) % |
|---|---|
| I want to have a laptop with an alternative keyboard layout. | 8 |
| I think it is necessary to type on alternative keyboard layouts. | 41 |
| I have no prejudices regarding alternative keyboard layouts. | 66 |
| I believe typing fast on alternative keyboard layouts will contribute to my professional career. | 66 |
| I want to type with ten fingers on alternative keyboard layouts. | 66 |
| I believe typing fast on alternative keyboard layouts is possible. | 92 |
| Within five years, it is possible for me to learn typing fast on alternative keyboard layouts. | 50 |

## 5. *Discussion*

Ergonomists often confront with challenges while introducing ergonomic interventions due to the negative impact that may affect workers' short-term productivity [17]. Although this negative impact reduces short-term productivity, multiple accomplishment of tasks will contribute to experience and increase productivity rates. In the case of an innovative intervention, exceeding the former productivity levels after a while will not be a surprise.

There is a wide range of research on the correlation between keyboard usage and health risks such as Carpal Tunnel Syndrome, Upper Extremity Disorders etc. Posture, physical shape of keyboard (angled, split etc.), height of keyboard and some other parameters contribute to these health risks [21], but we could not find any satisfying study comparing the Turkish keyboard layouts in terms of health risks during our literature review.

The learning percentages of participants prove that multiple trials have improved the participants' completion time. A comparison of the learning percentages from protocol-1 and protocol-2 indicates that the number of trials, 5 or 10, has no significant effect on the learning percentage, which is completely consistent with a previous study in which the findings from 5 and 20 trials were compared and no significant difference was observed [17].

A comparison of the completion times for trials shows that the TR-F keyboard has a significant effect in terms of typing speed. Since no typing exercise was done nor instruction provided before our experiment, this significant effect cannot be generalized to general users. A comparison of performance on keyboards after a period of typing exercise is not within the scope of this study, but it can be studied in further research. In addition, it can be stated that typing fast on the TR-Q does not engender significant improvement in the typing performance on an alternative keyboard layout.

During alternative keyboard trials, all participants struggled to find the exact location of characters on keyboards. All the keyboards tested were in the same shape, colour and design. The arrangements of vowels on different keyboards were similar, and no extreme physical demand was expected by us when we designed the experimental methodology. From this point of view, the weak correlation between the physical demand and learning percentage is not a surprise. As all the keyboards were in the same shape, participants had to memorize where the characters were, which contributed to the correlation between the cognitive demand and learning percentage. This contribution may be the reason why the cognitive demand showed stronger correlation than the physical demand. The correlation between the perceptual demand and learning percentage is the strongest of all, which is a result of new layouts' compelling the participants to search the keyboards to find the characters they had not memorized.

The TR-E keyboard layout has higher demand rates compared to the TR-F in all the three categories, but only the physical demand rate has a significant difference. Given the longer completion times achieved during the TR-E trials, the participants might have rated the TR-E as physically more demanding than the TR-F.

According to the pre-experiment survey results, none of the participants needed a laptop with an alternative keyboard layout. Most of the participants thought that it was not necessary to type on alternative keyboard layouts and only half of them believed that typing fast on alternative keyboard layouts would contribute to their professional career. Most of the participants had prejudice regarding alternative keyboards, and if they do not work for a governmental body, they will not have to use alternative keyboard layouts – the government officially obliges public agents to type on the TR-F keyboard.

The post-experiment survey results prove that if individuals are given the opportunity to type on an alternative keyboard, their prejudice may be broken down and more users may spend time to type on alternative keyboards.

The participants have different attitudes toward alternative keyboard layouts and several possible reasons may account for these attitude differences. Since the TR-E keyboard is novel, possible reasons are presented considering the TR-F.

A study on students of Bureau Management and Manager Assistance Associate Degree Program found out that 51.4% of students regularly used the TR-F and 36.7% prefer the TR-F over the TR-Q [22]. Since fast typing is a prerequisite skill for these students and they expected to be employed in a governmental institution in the future, the high percentage of TR-F keyboard preference is not fortuitous. On the other hand, our participants who study computer engineering may not have the same motive for typing on the TR-F keyboard.

Only 42.7% of the participants attending our study mainly type in Turkish. Since the TR-F keyboard is designed for Turkish, improving their typing skills on the TR-F would not dramatically contribute to the typing performance of those who mainly type in English (58.3%) in their daily life.

A great number of the participants attending our study spend most of their time on computer playing computer games and most of the computer games do not support built-in pre-assigned hotkey modification for Turkish alternative keyboard layouts. Many programming and compiling software has the same deficiency as well.

Lastly, some noteworthy points were revisited during our research regarding laptops with TR-F keyboards. In the Turkish market, the sales volume and range of laptop models with TR-F keyboard are limited (only three models have been found available in stock on e-commerce websites) and popular e-commerce websites still do not have "filter by keyboard" option for searching laptops. Yet, stickers for modifying keyboards are common and users buy TR-Q laptops and modify their keyboards into TR-F by means of a sticker. Likewise, plugging a second (TR-F) keyboard to laptops via USB is also common. If a user is used to typing on the TR-F, then they will need to carry their computer/keyboard with them or type on a QWERTY when they are abroad since the TR-F keyboard is not widespread. All these facts indicate a lack of coordination between the government and suppliers in terms of the government's effort to boost the popularity of TR-F keyboard.

## 6. Conclusion

In this study, we analyse and compare the Turkish-F and Turkish-E keyboard layouts through a user-based testing. The most important results pertaining to the experimental and survey data collected from our 12-participant user experiment are as follows:

- The learning percentage for the Turkish-F keyboard is slightly higher than that for the Turkish-E keyboard, but no significant difference has been observed.
- The completion time with the Turkish-F keyboard is significantly less than that with the Turkish-E keyboard.
- The Turkish-E keyboard has higher demand rates than the Turkish-F keyboard in all the three categories: cognitive, physical and perceptual demands. However, there is a statistical difference only for the physical demand rate.
- 66.7% of our subjects reported that their main activities on computer included gaming. Thus, it is not surprising to observe that only 16% of them find it necessary to type with a keyboard layout alternative to QWERTY. However, we are surprised to see that this percentage has increased to 41% after having a chance to try one of these alternatives.

The most important limitation for our work is the limited size of our participant pool. In addition, the freshman students majoring in computer engineering may not be representative of the general public. It can be a promising future study to conduct similar user experiments to understand if and how the results would change with increased diversity and size of users.